\documentclass[useAMS,usenatbib]{mn2e}

\usepackage{natbib}
\usepackage{amsmath,amsfonts,amssymb}
\usepackage[dvips]{graphicx}
\usepackage{longtable}

\title[Variability of \textit{WMAP} point sources]
  {Follow-up observations at 16 and 33\:GHz of 
   extragalactic sources from \textit{WMAP} 3-year data: II -- Flux-density variability
   \thanks{We request that any reference to this 
   paper cites `AMI Consortium: Franzen et al. 2009'}}
\author[AMI Consortium: Franzen et al.]
  {AMI Consortium:
  Thomas~M.~O.~Franzen,$^1$\thanks{Email: t.franzen@mrao.cam.ac.uk}
  Matthew~L.~Davies,$^1$\thanks{Email: m.davies@mrao.cam.ac.uk}
  \newauthor
  Rod~D.~Davies,$^2$
  Richard~J.~Davis,$^2$
  Farhan~Feroz,$^1$
  \newauthor
  Ricardo~G\'{e}nova-Santos,$^{1,3}$
  Keith~J.~B.~Grainge,$^{1,4}$
  David~A.~Green,$^1$
  \newauthor  
  Michael~P.~Hobson,$^1$
  Natasha~Hurley-Walker,$^1$
  Anthony~N.~Lasenby,$^{1,4}$
  \newauthor  
  Marcos~L\'{o}pez-Caniego,$^1$
  Malak~Olamaie,$^1$
  Carmen~P.~Padilla-Torres,$^3$
  \newauthor
  Guy~G.~Pooley,$^1$
  Rafael~Rebolo,$^3$
  Carmen Rodr\'{i}guez-Gonz\'{a}lvez,$^1$
  \newauthor  
  Richard~D.~E.~Saunders,$^{1,4}$
  Anna~M.~M.~Scaife,$^1$
  Paul~F.~Scott,$^1$
  \newauthor   
  Timothy~W.~Shimwell,$^1$
  David~J.~Titterington,$^1$
  Elizabeth~M.~Waldram,$^1$
  \newauthor 
  Robert~A.~Watson$^2$
  and Jonathan~T.~L.~Zwart$^1$\\
  $^1$Astrophysics Group, Cavendish Laboratory,
      19 J.~J.~Thomson Avenue, Cambridge CB3 0HE \\
  $^2$Jodrell Bank Centre for Astrophysics,
      University of Manchester, Manchester M13 9PL\\
  $^3$Instituto de Astrof\'{i}sica de Canarias,
      38200 La Laguna, Tenerife, Canary Islands,
      Spain\\
  $^4$Kavli Institute for Cosmology Cambridge,
      Madingley Road, Cambridge, CB3 0HA}
\date{Accepted ????. Received ????}
\pagerange{\pageref{firstpage}--\pageref{lastpage}}
\pubyear{2009}

\newcommand{\vect}[1]{\mathbf{#1}} 

\begin{document}
\maketitle
\label{firstpage}
\begin{abstract}

Using the Arcminute Microkelvin Imager (AMI) at 16~GHz and the Very Small Array (VSA) at 33~GHz to make follow-up observations of sources in the New Extragalactic \textit{WMAP} Point Source Catalogue, we have investigated the flux-density variability in a complete sample of 97 sources over timescales of a few months to $\approx 1.5$~years.

We find that 53 per cent of the 93 sources, for which we have multiple observations, are variable, at the 99~per~cent confidence level, above the flux density calibration uncertainties of $\approx 4$~per~cent at 16~GHz; the fraction of sources having varied by more than 20 per cent is 15 per cent at 16~GHz and 20 per cent at 33\:GHz. Not only is this common occurrence of variability at high frequency of interest for source physics, but strategies for coping with source contamination in CMB work must take this variability into account.

There is no strong evidence of a correlation between variability and flux density for the sample as a whole. For those sources classified as variable, the mean fractional r.m.s. variation in flux density increases significantly with the length of time separating observation pairs.  Using a maximum-likelihood method, we calculate the correlation in the variability at the two frequencies in a subset of sources classified as variable from both the AMI and VSA data and find the Pearson product-moment correlation coefficient to be very high ($0.955 \pm 0.034$). We also find the degree of variability at 16~GHz ($0.202 \pm 0.028$) to be very similar to that at 33\:GHz ($0.224 \pm 0.039$).

Finally, we have investigated the relationship between variability and spectral index, $\alpha_{13.9}^{33.75}$ (where $S\propto \nu^{-\alpha}$), and find a significant difference in the spectral indices of the variable sources ($-0.06 \pm 0.05$) and non-variable sources ($0.13 \pm 0.04$). 

\end{abstract}
\nokeywords

\section{Introduction}\label{Introduction}

The existence of the time variability of the flux densities of some extragalactic radio sources has been established since the 1960s \citep[see e.g.][]{dent}. Long-standing issues, stemming mostly from observations at $\lesssim 5$~GHz, have concerned the effects of the path between the source and us -- with implications for the propagation medium -- and of the possible roles of variability of the source luminosity -- with implications for how the sources work. More recently, as studies of structure in the CMB have become possible (requiring observations at $\gtrsim 10$~GHz), knowledge of variability at high frequency is becoming essential in order to remove CMB foreground contamination. We provide some illustration of the importance and causes of variability, as follows.

\subsubsection{Interstellar scintillation}

\citet{shapirovskaya} and \citet{rickett1984} suggested that refractive interstellar scintillation (ISS) could be responsible for the intra-day variability of extragalactic radio sources at $\nu \lesssim 5$~GHz.  The basis of the mechanism is that the refractive index, $n$, from an element of plasma in a cloud in the interstellar medium (ISM) in our galaxy has the usual form $n = [1-(\frac{\nu_{p}}{\nu})^{2}]^{1/2}$, where $\nu$ is the observation frequency and $\nu_{\mathrm{p}}$ is the plasma frequency, but that the variation in n in the plane of the sky across the cloud (or indeed resulting from more than one cloud in the line of sight) produces scintillation as the galactic ISM moves with respect to both us and the radio source. The form of $n(\nu)$ indicates that these scintillations will not be important at high enough $\mathrm{\nu}$.

Many studies \citep[see e.g.][]{gregorini,cawthorne,spangler} support this notion in that they show that variability tends to increase as the galactic latitude $\vert b \vert$ of the sources decreases. Further, \citet{dennett-thorpe}, following a technique due to \citet{lang} and \citet{jauncey}, used two telescopes at different locations on Earth to demonstrate the effect on the flux density at 4.9\:GHz of J1819+3845 of the Earth moving through a scintillation pattern at the Earth caused by the local ISM. Finally, \citet{linsky} have shown unequivocally that the motion of individual clouds in the local ISM, especially as cloud edges pass the line of sight, is an important mechanism of variability at 5\:GHz. 

\subsubsection{Intrinsic source variability}

It is also clear that there is a component of variability that is intrinsic to some sources. The existence of this type of variability is fully supported by VLBI observations of the motion and luminosity evolution of synchrotron emitting components \citep[see e.g.][]{zensus}. For a moving source, $S_{\mathrm{ob}}$, the observed flux density, is related to $S$, the flux density that would be observed at the same frequency $\nu$ in the comoving frame, by $S_{\mathrm{ob}}(\nu) = S(\nu)D^{3+\alpha}$, where $\alpha$ is the spectral index and $D$ the boost factor which depends on both the speed of the source and its angle to the line of sight. (This equation is true for optically thin sources and spherical optically thick sources.) For a relativistically moving source, curvature in the trajectory can therefore produce flux density variation \citep[see e.g.][]{blandford}.

\citet{rickett2006} have discussed the degree of variability as a function of $\nu$. They studied 146 extragalactic compact radio sources (a very important but not a complete sample of all sources above a given flux density) monitored over 1979--1996 with the Green Bank Interferometer at 2 and 8\:GHz; major conclusions are that the ISS component of variability is much stronger at 2\:GHz than 8\:GHz, and that intrinsic variability is of prime importance at high radio frequency.

We now focus on variability at $\mathrm{\geq 15\:GHz}$. Given the foregoing work, we expect this variability to be dominated by intrinsic luminosity variability -- so at $\mathrm{\geq 15\:GHz}$ we are dealing with the intrinsic properties of the radio sources -- although the work of \citeauthor{linsky} clearly implies that a small fraction of sources will have substantial variability caused by the ISS as the edges of ISM clouds pass the line of sight between us and the radio source, even at high $\nu$.

\subsubsection{Our work}

There are two large-scale surveys of extragalactic radio sources that are complete in flux density at $\mathrm{\geq 15\:GHz}$ and which study variability: the 9th Cambridge (9C) survey complete to $\approx$~25\:mJy at 15\:GHz \citep{bolton} and the Australia Telescope Compact Array (ATCA) 20-GHz pilot survey complete to $\approx$~100\:mJy \citep{sadler}. Here, we study variability in a high-flux-density ($\mathrm{>1~Jy}$), very-large-area survey. As well as being complementary to the two other studies, our work has implications for forthcoming CMB work.

Our work follows on from \citet{davies} (Paper I) which employs the New Extragalactic \textit{WMAP} Point Source (NEWPS) catalogue and is complete to 1.1\:Jy at 33\:GHz. Using the Arcminute Microkelvin Imager \citep[AMI;][]{zwart} at 16\:GHz and the Very Small Array \citep[VSA;][]{watson} at 33\:GHz, we have carried out simultaneous observations of a subset of the sample to investigate variability over timescales ranging from a few months to $\mathrm{\approx}$ 1.5~years.  All the AMI data presented in this paper were obtained using the AMI Small Array.  Observation and reduction strategies are as described in Paper I.

\section{Flux Density Variability}\label{Flux Variability}

\subsection{Variability of the source population as a whole}

\subsubsection{Previous work}

\citet{bolton} studied the variability of 51 faint sources from the 9C survey (complete to $\approx$~25\:mJy at 15\:GHz) at 15\:GHz on a 3-year timescale. They defined the fractional variation as 
\begin{eqnarray}
\label{equation:var_index_bolton}
\frac {\Delta X}{X} = \frac{1}{\mu} \sqrt{\frac{1}{n}\displaystyle\sum_{i=1}^n (x_{i} - \mu)^{2}}
\mathrm{,}
\end{eqnarray}
where $x_{i}$ are individual flux density measurements for the same source, $n$ is the number of data points and $\mu$ is the mean flux density. For the 17 sources in \citeauthor{bolton} classified as variable, the fractional variation ranges between 8.4 and 70~per~cent, with a median of 14~per~cent.

\citet{sadler} studied the variability of 108 sources from the ATCA 20-GHz pilot survey (complete to 100\:mJy at 20\:GHz) at 20\:GHz on a 1--2~year timescale. For this purpose, they used follow-up observations at two epochs during 2003--04. They quantified the variability in the same way as \citeauthor{bolton}, except that they took the uncertainties in individual flux density measurements into account. They defined the unbiased variability index, $V_{\mathrm{rms}}$ by
\begin{eqnarray}
\label{equation:var_index_sadler}
V_{\mathrm{rms}} = \frac{1}{\mu} 
\sqrt{\frac{1}{n}\left[\displaystyle\sum_{i=1}^n (x_{i} - \mu)^{2} - \displaystyle\sum_{i=1}^n \sigma_{i}^{2}\right]}
\mathrm{,} 
\end{eqnarray}
where $\sigma_{i}$ are errors on individual flux density measurements. This expression follows \citet{akritas}, who make it clear that the value of $\mathrm{\sigma_{i}}$ must be the same for each of the observations of a particular source. For all sources, \citeauthor{sadler} obtained a median value of $V_{\mathrm{rms}}$ at 20\:GHz of 6.9 per cent. Table~\ref{tab:var_index_sadler} shows their distribution of $V_{\mathrm{rms}}$.

\subsubsection{Methods}

Good knowledge of the uncertainties in flux density measurements is crucial for the study of variability. Great care was put into determining these before taking them into account in quantifying the variability (see Paper I). Because the VSA is not equatorially mounted, the linear polarisation measured with the VSA changes as a function of parallactic angle. However, we almost entirely overcome this difficulty by observing the sources at very similar HAs in different observing runs.  We also note that, if polarisation were playing an important role, we would have expected to see greater scatter in the measured flux densities for each source within a single observing run.  Moreover, in general the \textit{WMAP}-detected point sources are not strongly polarised: for the sources detected in the \textit{WMAP} 5-year maps, \citet{wright} have measured a mean polarisation at 33\:GHz of just 2.2 per cent, although we have not determined whether there is a relation between polarisation and variability.  We therefore do not expect polarisation to be an important factor in the variability observed with the VSA.

\begin{table}
 \caption{Distribution of the unbiased variability index $V_{\mathrm{rms}}$ at 20\:GHz for sources in the \citet{sadler} sample.}
 \label{tab:var_index_sadler}
  \begin{tabular}{c c c }
 \hline
 $V_{\mathrm{rms}}$ (per cent) & $n$ & Fraction (per cent) \\                      
 \hline
 $ <10    $ &  63 & $ 58 \pm 7 $ \\
   10--20   &  29 & $ 27 \pm 5 $ \\
   20--30   &  11 & $ 10 \pm 3 $ \\
 $ >30    $ &   5 & $  5 \pm 2 $ \\
 Total      & 108 \\
 \hline
 \end{tabular}
\end{table} 

\begin{table}
 \caption{Distribution of the unbiased variability index $V$ at 16 and 33\:GHz for sources in our sample.  The errors on the fractions are Poisson errors.}
 \label{tab:var_index}
 \begin{tabular}{c c c c c }
 \hline
 $V$ (per cent) & \multicolumn{2}{c}{$n$} & \multicolumn{2}{c}{Fraction (per cent)} \\
                      & 16~GHz & 33~GHz & 16~GHz & 33\:GHz \\
 \hline
 $ <10    $ & 61 & 55 & $ 66 \pm 8 $ & $ 59 \pm  8 $ \\
   10--20   & 18 & 19 & $ 19 \pm 5 $ & $ 20 \pm  5 $ \\
   20--30   &  8 & 12 & $  9 \pm 3 $ & $ 13 \pm  4 $ \\
 $ >30    $ &  6 &  7 & $  7 \pm 3 $ & $  8 \pm  3 $ \\
 Total      & 93 & 93 \\
 \hline
 \end{tabular}
\end{table}

\begin{figure}
\includegraphics[scale=0.6,angle=0]{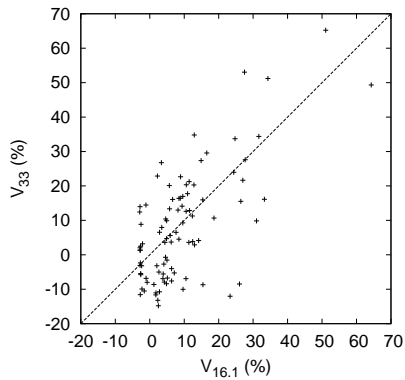}
\caption{Unbiased variability index $V$ at 16 against that at 33\:GHz.}
\label{fig:var_index}
\end{figure}

We have tried to analyse the variability in such a way as to make our results comparable to those obtained by \citet{sadler}. So, we have only made use of the first and last flux density measurements from both telescopes for each source, yielding similar time intervals. In the AMI data, time intervals range between 1 and 18 months, with a median time interval of 14 months, and in the VSA data, time intervals range between 2 and 17 months, with a median time interval of 16 months. We have not quantified the variability using Eqn.~\ref{equation:var_index_sadler}, as it is not suitable in the case where the level of variability is less than approximately the uncertainties on the flux density measurements and the number of data points is small; this is because it does not take account of the error on the mean flux density.

Instead, we define the unbiased variability index for a pair of flux density measurements as follows: let $S_{1}$ and $S_{2}$ be the two measured flux densities with errors, including contributions from both calibration and thermal noise, $\sigma_{1}$ and $\sigma_{2}$ respectively. We assume that, on average, the measured change in flux density is related to the true change in flux density, $\sigma_{\mathrm{var}}$, by  
\begin{eqnarray}
\label{equation:var_index_der1}
(S_{1}-S_{2})^{2} = \sigma_{\mathrm{ins}}^{2} + \sigma_{\mathrm{var}}^{2}
\mathrm{,}
\end{eqnarray}
where $\sigma_{\mathrm{ins}} = \sqrt{\sigma_{1}^2 + \sigma_{2}^2}$ is the measurement error on $(S_{1}-S_{2})$. The true change in flux density can therefore be expressed as
\begin{eqnarray}
\label{equation:var_index_der2}
\sigma_{\mathrm{var}} = \sqrt{(S_{1}-S_{2})^{2} - \sigma_{\mathrm{ins}}^{2}}
\mathrm{.}
\end{eqnarray}

We then define our unbiased variability index for a pair of flux density measurements as
\begin{eqnarray}
\label{equation:var_index}
V = \frac{1}{2 \bar{S}} 
\sqrt{(S_{1} - S_{2})^{2} 
- (\sigma^{2}_{1} + \sigma^{2}_{2})} 
\mathrm{,}
\end{eqnarray}
where $\bar{S}$ is the mean flux density. The factor $1/2\bar{S}$ ensures that $V$ lies between 0 and 1, as is the case with $V_{\mathrm{rms}}$.  $V$ converges to $V_{\mathrm{rms}}$ when the level of variability becomes large compared with the uncertainties on the flux density measurements.

When the value inside the square root in Eqn.~\ref{equation:var_index} becomes negative, following a method similar to \citet{barvainis}, we set the unbiased variability index to be negative and this is given by
\begin{eqnarray}
\label{equation:var_index2}
V = - \frac{1}{2 \bar{S}} 
\sqrt{\vert{(S_{1} - S_{2})^{2} 
- (\sigma^{2}_{1} + \sigma^{2}_{2})}\vert} 
\mathrm{.}
\end{eqnarray}

\subsubsection{Results}

Figure~\ref{fig:var_index} plots our unbiased variability index at 33~GHz against that at 16~GHz.  It illustrates that the levels of variability, for sources in this sample, are broadly correlated at the two frequencies.  However, the plot cannot be used for detailed comparison of the variability at the two frequencies, because, for individual sources, the observations were carried out with different time intervals using the two telescopes.

The median values of $V$ at 16 and 33~GHz are 6.2 and 4.7 per cent respectively. We stress that the median value of $V$ at 33\:GHz is unreliable because variability detected at this level with the VSA is not significant. Table~\ref{tab:var_index} shows the distribution of $V$ at 16 and 33~GHz for the sources in our sample.

The most variable source from the AMI data is J1419+3822 with $V = 64$~per~cent on a timescale of 524 days, the source having decreased in flux density by about a factor of five; the corresponding value of $V$ from the VSA data is 49~per~cent. The most variable source from the VSA data is J1727+4530 with $V = 65$~per~cent on a timescale of 486 days, the source having increased in flux density by about a factor of five; the corresponding value of $V$ from the AMI data is 51~per~cent.

We find a similar level of variability in our source sample to that found by \citet{sadler}.  This suggests that the two samples have very similar source properties, despite the difference of an order of magnitude in flux density levels.

Given the level of variability that we find, over a significantly shorter time period than separates our observations from those carried out by \textit{WMAP}, it seems likely that the Eddington bias, found in the comparison of AMI/VSA and \textit{WMAP} flux densities discussed in Paper~I, can resonably be attributed to variability.

\subsection{Source classification}

Following \citet{bolton}, we have used a $\mathrm{\chi^{2}}$-test to classify each of the sources as variable (VAR) or non-variable (NV) at both our observing frequencies.  Our null hypothesis is that the source has a constant flux density and that the $i$th measurement of the source's flux density, $x_{i}$, is drawn from a Gaussian distribution with a mean, $\mu$ (which we take as the mean flux density measured for that source), and a standard deviation, $\sigma_{i}$, given by the uncertainty on the $i$th measurement of the flux density.
The value of $\chi^{2}$ is given, therefore, by
\begin{eqnarray}
\chi^{2} =  \displaystyle\sum_{i=1}^n \frac{(x_{i} - \mu)^{2}}{\sigma^2_{i}}
\mathrm{,}
\end{eqnarray}
where $n$ is number of observations made of the source.  When the degree of confidence that the data do not support the null hypothesis is greater than 99 per cent we define the source as variable, otherwise we accept the null hypothesis and the source is defined as non-variable.  We have included the classification for each source at both observing frequencies in Table~\ref{tab:sources}.  The mean separation between all pairs of observations for each of the sources is also included.  We stress that, since this value varies from source to source, the timescale on which we sample variability also varies with source.

At 16~GHz, of the 93 sources, for which we have more than a single observation, we classify 49 (53 per cent) as variable.  At 33\:GHz, we have more than one observation for 96 of the 97 sources in the sample; of these we classify 30 (31 per cent) as variable.  We note that if we use a 90~per~cent confidence level in the chi-squared test, the percentages of sources classified as variable are 65 and 49 at 16 and 33~GHz respectively.  Of the 30 sources defined as variable at 33\:GHz, 26 are also classified as variable at 16~GHz.  This leaves four sources which are only classified as variable at 33\:GHz; however, we believe that this is because the observations of these four sources spanned somewhat different date ranges on the two telescopes (see Fig.~\ref{fig:var_sources}).  The higher proportion of sources classified as variable at the lower frequency is, in large part, due to the larger errors on the flux densities measured by the VSA.  We attempt, below, to make an assessment of the correlation in the variability between the two frequencies for those sources defined as variable at both 16 and 33\:GHz, taking account of the measurement errors. In Fig.~\ref{fig:var_sources} we have provided the light curves of these sources.

\begin{figure}
\includegraphics[scale=0.35,angle=270]{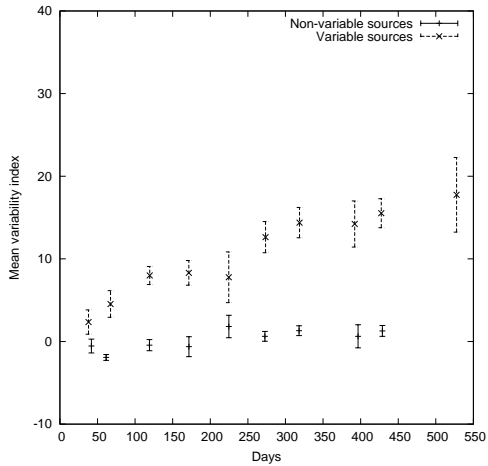}
\caption{The mean variability index at 16~GHz versus number of days separating the observation pairs for those sources classified as variable from the AMI data (dashed) and non-variable from the AMI data (solid).  The data have been averaged using bins of 50 days in width.}
\label{fig:days_vs_variation_ami}
\end{figure}

\begin{figure}
\includegraphics[scale=0.35,angle=270]{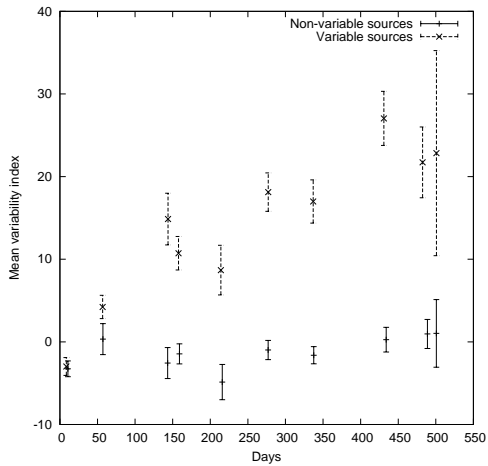}
\caption{The mean variability index at 33\:GHz versus number of days separating the observation pairs for those sources classified as variable from the VSA data (dashed) and non-variable from the VSA data (solid).  The data have been averaged using bins of 50 days in width.}
\label{fig:days_vs_variation_vsa}
\end{figure}

We have attempted to assess the variability as a function of time, at both observing frequencies, for those sources which we have classified as variable.  We have calculated the variability index (equations \ref{equation:var_index} and \ref{equation:var_index2}) for every pair of observations for these sources.  Figures~\ref{fig:days_vs_variation_ami} and \ref{fig:days_vs_variation_vsa} show plots of the variability index versus number of days separating the observation pairs at 16 and 33~GHz, respectively.  The data have been averaged within bins of 50 days in width; the mean variability index is shown for observation pairs falling within each of the bins, along with the standard error on the mean.  We have also carried out an identical procedure for those sources classified as non-variable; the results have been included in the figures for comparison.

The plots indicate that, at both frequencies, the mean variability index for those sources defined as variable, increases significantly with time.  In contrast, for the non-variable sources, there is little change in the mean variability index with time, which, in both cases, are close to zero.  The mean variability index is generally higher for the variable sources at 33\:GHz compared with that at 16~GHz.  However, due to the larger measurement errors on the VSA observations, only the most variable sources are selected as variable at 33\:GHz; for the same reason, the level of variability required for a source to be classified as variable is likely to be greater at 33 than 16~GHz.  This could explain why the mean variability is generally greater at the higher observing frequency.

\begin{figure}
\includegraphics[scale=0.35,angle=270]{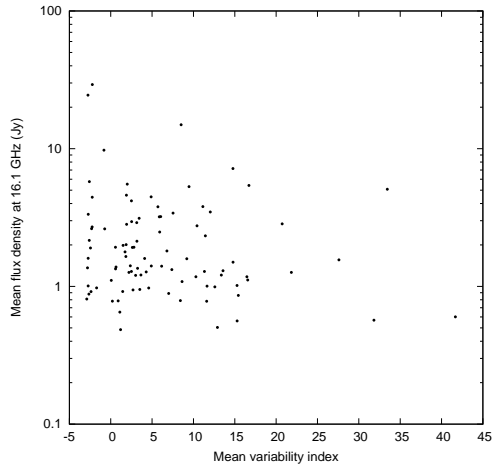}
\caption{Mean variability index versus mean flux density at 16\:GHz.}
\label{fig:mean_flux_vs_mean_var_index_ami}
\end{figure}

\begin{figure}
\includegraphics[scale=0.35,angle=270]{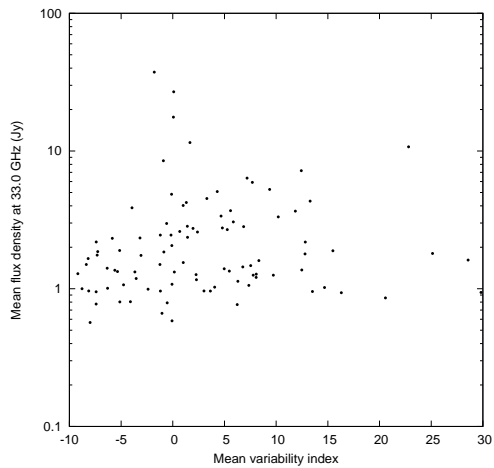}
\caption{Mean variability index versus mean flux density at 33\:GHz.}
\label{fig:mean_flux_vs_mean_var_index_vsa}
\end{figure}

Figure~\ref{fig:mean_flux_vs_mean_var_index_ami} shows the mean variability index versus mean flux density at 16\:GHz, for sources that were observed multiple times at that frequency.  There appears to be no strong correlation between flux density and variability.  We compared the mean variability index of the brightest 50~per~cent of sources to the faintest 50~per~cent; we found no statistically significant difference in the mean variability index of the brighter sample compared to the fainter.  Figure~\ref{fig:mean_flux_vs_mean_var_index_vsa} shows a similar plot at 33\:GHz.  There are a number of fainter sources with low variability indices; this is, however, to be expected due to the significant contribution of thermal noise to the measurement uncertainties for the fainter sources at 33\:GHz.  Again, we found no statistically significant difference in the mean variability between the brighter and fainter sources.

\begin{figure*}
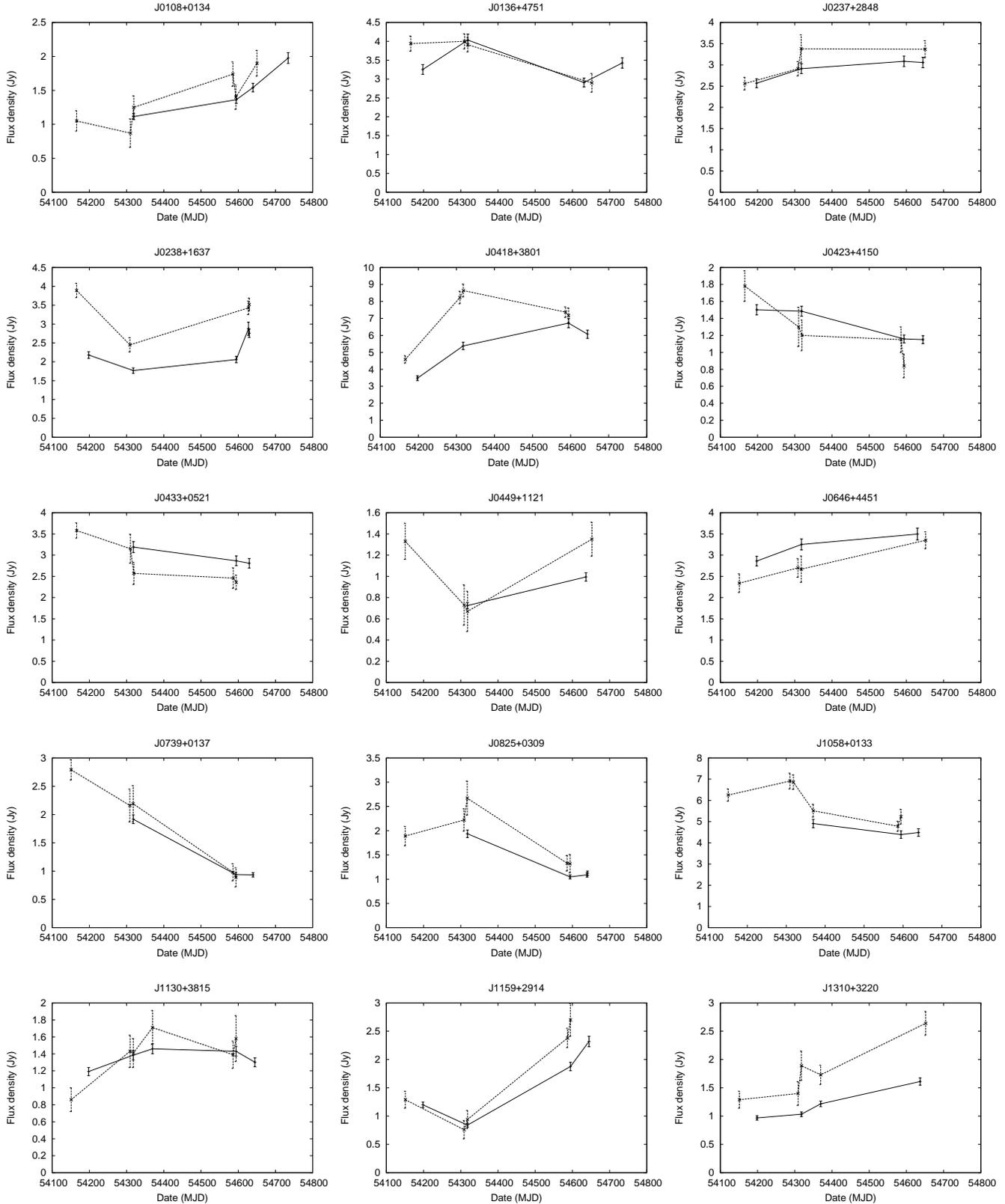

  \label{fig:var_sources}   
  \includegraphics[scale=0.46,bb=0.7in 0.6in 5.7in 4.3in,clip=]{./figures/J0108+0134.epsi}
  \includegraphics[scale=0.46,bb=0.7in 0.6in 5.7in 4.3in,clip=]{./figures/J0136+4751.epsi}
  \includegraphics[scale=0.46,bb=0.7in 0.6in 5.7in 4.3in,clip=]{./figures/J0237+2848.epsi} \\ 
  \includegraphics[scale=0.46,bb=0.7in 0.6in 5.7in 4.3in,clip=]{./figures/J0238+1637.epsi}
  \includegraphics[scale=0.46,bb=0.7in 0.6in 5.7in 4.3in,clip=]{./figures/J0418+3801.epsi}
  \includegraphics[scale=0.46,bb=0.7in 0.6in 5.7in 4.3in,clip=]{./figures/J0423+4150.epsi} \\  
  \includegraphics[scale=0.46,bb=0.7in 0.6in 5.7in 4.3in,clip=]{./figures/J0433+0521.epsi}	      
  \includegraphics[scale=0.46,bb=0.7in 0.6in 5.7in 4.3in,clip=]{./figures/J0449+1121.epsi}  
  \includegraphics[scale=0.46,bb=0.7in 0.6in 5.7in 4.3in,clip=]{./figures/J0646+4451.epsi} \\	   
  \includegraphics[scale=0.46,bb=0.7in 0.6in 5.7in 4.3in,clip=]{./figures/J0739+0137.epsi}
  \includegraphics[scale=0.46,bb=0.7in 0.6in 5.7in 4.3in,clip=]{./figures/J0825+0309.epsi}  
  \includegraphics[scale=0.46,bb=0.7in 0.6in 5.7in 4.3in,clip=]{./figures/J1058+0133.epsi} \\	    
  \includegraphics[scale=0.46,bb=0.7in 0.6in 5.7in 4.3in,clip=]{./figures/J1130+3815.epsi}
  \includegraphics[scale=0.46,bb=0.7in 0.6in 5.7in 4.3in,clip=]{./figures/J1159+2914.epsi}
  \includegraphics[scale=0.46,bb=0.7in 0.6in 5.7in 4.3in,clip=]{./figures/J1310+3220.epsi} \\
  \caption{Flux densities at 16 (solid line) and 33\:GHz (dashed line) of the sources which are classified as variable from the VSA data. Four of these sources (J0433+0521, J1058+0133, J1608+1029 and J2148+0657) are not classified as variable from the AMI data but this is thought to be because the observations of these four sources spanned somewhat different date ranges on the two telescopes.}
\end{figure*}

\begin{figure*}
  \contcaption{}
  \includegraphics[scale=0.46,bb=0.7in 0.6in 5.7in 4.3in,clip=]{./figures/J1419+3822.epsi}  
  \includegraphics[scale=0.46,bb=0.7in 0.6in 5.7in 4.3in,clip=]{./figures/J1608+1029.epsi}	  
  \includegraphics[scale=0.46,bb=0.7in 0.6in 5.7in 4.3in,clip=]{./figures/J1642+3948.epsi} \\
  \includegraphics[scale=0.46,bb=0.7in 0.6in 5.7in 4.3in,clip=]{./figures/J1727+4530.epsi} 
  \includegraphics[scale=0.46,bb=0.7in 0.6in 5.7in 4.3in,clip=]{./figures/J1734+3857.epsi}
  \includegraphics[scale=0.46,bb=0.7in 0.6in 5.7in 4.3in,clip=]{./figures/J1740+5211.epsi} \\
  \includegraphics[scale=0.46,bb=0.7in 0.6in 5.7in 4.3in,clip=]{./figures/J1743-0350.epsi} 
  \includegraphics[scale=0.46,bb=0.7in 0.6in 5.7in 4.3in,clip=]{./figures/J1751+0938.epsi}  
  \includegraphics[scale=0.46,bb=0.7in 0.6in 5.7in 4.3in,clip=]{./figures/J2148+0657.epsi} \\	     
  \includegraphics[scale=0.46,bb=0.7in 0.6in 5.7in 4.3in,clip=]{./figures/J2202+4216.epsi}
  \includegraphics[scale=0.46,bb=0.7in 0.6in 5.7in 4.3in,clip=]{./figures/J2203+1725.epsi} 
  \includegraphics[scale=0.46,bb=0.7in 0.6in 5.7in 4.3in,clip=]{./figures/J2232+1143.epsi} \\
  \includegraphics[scale=0.46,bb=0.7in 0.6in 5.7in 4.3in,clip=]{./figures/J2236+2828.epsi} 
  \includegraphics[scale=0.46,bb=0.7in 0.6in 5.7in 4.3in,clip=]{./figures/J2253+1608.epsi}
  \includegraphics[scale=0.46,bb=0.7in 0.6in 5.7in 4.3in,clip=]{./figures/J2327+0940.epsi} \\  
\end{figure*}

\subsection{Correlation in the variability at 16 and 33\:GHz}

\subsubsection{Methods}

\begin{figure}
\includegraphics[scale=0.6,angle=0]{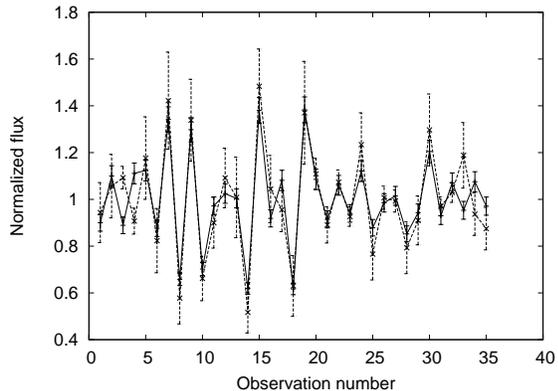}
\caption{Joint data-set, formed from pairs of simultaneous observations at 16~GHz (solid line) and 33\:GHz (dashed line) for 16 variable sources, for which the Pearson product-moment correlation coefficient is estimated.}
\label{fig:correlation_data}
\end{figure}

We consider the group of 26 sources which are classified as variable from both the AMI and VSA data.  For further analysis, we have required that a source has had at least two pairs of observations made simultaneously with the AMI and VSA; this produces a set of 16 sources. It was not possible to treat sources individually in this analysis, because, with only two or three pairs of simultaneous observations per source, the problem was too unconstrained. Instead, we treated the 16 sources collectively, so that, with a total of 35 pairs of simultaneous observations, the problem was well constrained. Figure \ref{fig:correlation_data} shows our joint dataset. Note that, before concatenating the data, we normalized the fluxes so that the mean AMI and VSA fluxes of each source are equal to 1.0. The range of time intervals between observations is 51--277 days and the average time interval is 162 days.

For this joint dataset, we then estimated the Pearson product-moment correlation coefficient $R$ using a maximum-likelihood method.  To carry this out, we form the likelihood of the data given covariance matrices for the signal and noise.  Let $x_{i}$ and $y_{i}$ be the $i^{th}$ measurements of the flux densities at 16 and 33\:GHz respectively, $\sigma_{x,i}$ and $\sigma_{y,i}$ the uncertainties on $x_{i}$ and $y_{i}$ respectively and $n$ the total number of measurements.  The log-likelihood function is given by
\begin{eqnarray}
\lefteqn{\mathrm{ln} L(\vect{q}) =} \\ & & 
\mathrm{constant} - \frac{1}{2} \displaystyle \sum_{i=1}^n \left[ \mathrm{ln} \vert \vect{C}_i \vert + (\vect{x}_i - \vect{m})^t \vect{C}^{-1}_{i}
(\vect{x}_i - \vect{m}) \right] \mathrm{,} 
\end{eqnarray}
where $(\vect{x}_i - \vect{m})^t$ is the transpose of 
$(\vect{x}_i - \vect{m})$,
\begin{eqnarray}
\vect{C}_i & = & \vect{S} + \vect{N}_i \\
& = & 
\left(
\begin{array}{cc}
S_{11} & S_{12} \\
S_{21} & S_{22}
\end{array}
\right)
+
\left(
\begin{array}{cc}
\sigma^{2}_{x,i} & 0 \\
0 & \sigma^{2}_{y,i} 
\end{array}
\right)
\mathrm{,}
\end{eqnarray}
\begin{eqnarray}
\vect{x}_i =
\left(
\begin{array}{c}
x_{i} \\
y_{i} 
\end{array}
\right) 
\mathrm{,\ }
\vect{m} =
\left(
\begin{array}{c}
m_{x} \\
m_{y} 
\end{array}
\right)
\mathrm{,} 
\end{eqnarray}
the vector $\vect{q} = (m_{x}, \quad m_{y}, \quad S_{11}, \quad S_{12}, \quad S_{22})$ represents the parameters of a Gaussian model to be determined by maximizing the likelihood, the parameters $m_{x}$ and $m_{y}$ are mean flux densitiess at 16 and 33\:GHz respectively, $S_{11}$ and $S_{22}$ are variances of flux densities at 16 and 33\:GHz respectively and $S_{12}$ ($= S_{21}$) is the covariance between flux densities at 16 and 33\:GHz.  Then, $R$ is a derived parameter and is given by 
\begin{equation}
R = \frac{S_{12}}{\sqrt{S_{11}S_{22}}} \nonumber
\mathrm{.} 
\end{equation}
We used the nested sampling algorithm as implemented in the MultiNest code \citep{feroz,feroz2} to sample points from $\mathrm{ln} (\vect{q})$ with the constraints $-1 \leq R \leq 1$ and $\mathrm{\vert \vect{C}_i \vert \geq 0}$.

\subsubsection{Results}

\begin{figure}
\includegraphics[scale=0.6,angle=0]{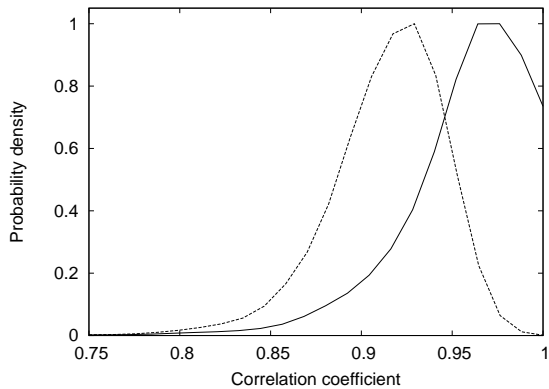}
\caption{The one-dimensional marginalized probability distribution for the Pearson product-moment correlation coefficient with (solid line) and without (dashed line) use of the noise covariance matrix term.}
\label{fig:correlation_mcmc}
\end{figure}

We obtained a value of $R = 0.955 \pm 0.034$ with a maximum likelihood value of 0.967.  We repeated the calculation setting the noise covariance matrix term to zero.  This yielded a value of $R = 0.914 \pm 0.030$, i.e. instrumental noise leads to a significant decrease in the degree of correlation unless account is taken of the noise.  This demonstrates the important role of the noise covariance matrix term in removing the bias in $R$ resulting from instrumental noise.  Figure~\ref{fig:correlation_mcmc} shows one-dimensional marginalized probability distributions for $R$ with and without use of the noise covariance matrix term. 

\begin{figure}
\includegraphics[scale=0.6,angle=0]{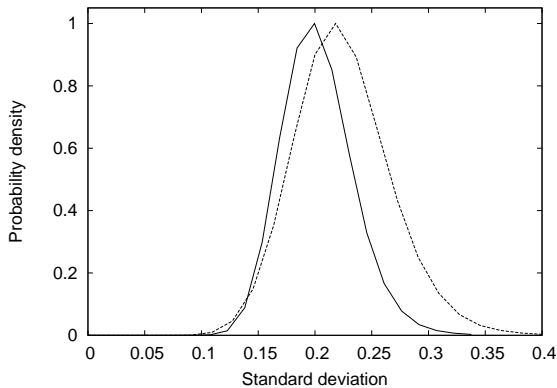}
\caption{The one-dimensional marginalized probability distribution for the standard deviations of flux densities at 16 (solid line) and 33\:GHz (dashed line)}.
\label{fig:sigma_mcmc}
\end{figure}

Our analysis has shown that, for the variable sources in general, over time scales of several months, flux densities at 16 and 33~GHz are very highly correlated.  Note that we have assumed that flux densities at 16 and 33~GHz are drawn from Gaussian distributions.  This is unlikely to be the case for individual sources observed over long periods of time.  However, in this particular case, we think it is a reasonable assumption to make given the fact that the analysis is run on a group of 16 sources with each source having approximately the same number of observations.  We have assumed that measurement errors at the two frequencies are Gaussian distributed and completely uncorrelated.  The latter is certainly true, since we are dealing with two telescopes at different sites.  Owing to the lack of data, it was not possible to treat sources individually, making it hard to ensure that sources are given equal weight.  We therefore stress that our results are not necessarily applicable to any individual source but rather to the population.

From this joint dataset, it is also possible to estimate the standard deviations of flux densities at 16 and 33~GHz, $s_{\mathrm{16}}$ and $s_{\mathrm{33}}$ respectively, using this method. Again, these are derived parameters and are given by $s_{\mathrm{16}} = \sqrt{S_{\mathrm{11}}}$ and $s_{\mathrm{33}} = \sqrt{S_{\mathrm{22}}}$. We obtained a value of $s_{\mathrm{16}} = 0.202 \pm 0.028$ with a maximum likelihood value of 0.188 and a value of $s_{\mathrm{33}} = 0.224 \pm 0.039$ with a maximum likelihood value of 0.198.  Figure~\ref{fig:sigma_mcmc} shows one-dimensional marginalized probability distributions for $s_{\mathrm{16}}$ and $s_{\mathrm{33}}$. 

These results suggest that there is little difference in the levels of variability at the two frequencies.  Previous studies suggest that there is only a slight increase in the degree of variability of compact radio sources with frequency.  The ATCA was used by \citet{tingay} to observe a sample of 202 sources from the VSOP all-sky survey \citep{hirabayashi} at 5\:GHz over a period of 3--4 years. They measured a median variability at 1.4\:GHz of 6 per cent and at 8.6\:GHz of 9 per cent.  \citet{owen} studied a group of sources with $S_{\mathrm{90~GHz}} > 1$~Jy and found only slightly more variability at 90\:GHz than at\:5 GHz.


\subsection{Relationship between variability and spectral properties}

\begin{figure}
\includegraphics[scale=0.35,angle=270]{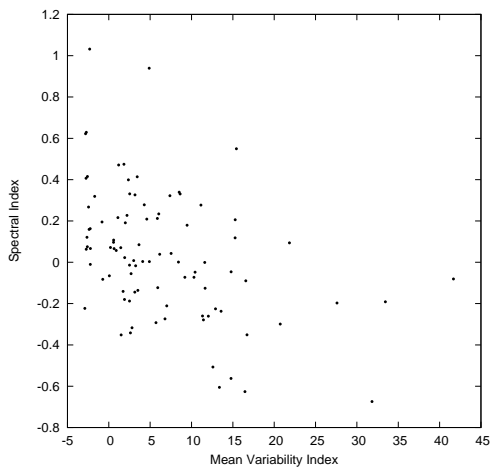}
\caption{Mean spectral index versus mean variability index at 16~GHz.}
\label{fig:spec_index_vs_var}
\end{figure}

We measured spectral indices between 13.9 and 33.75~GHz using simultaneous observations with the AMI and VSA (Paper I). Figure~\ref{fig:spec_index_vs_var} shows a plot of mean spectral index versus mean variability index at 16~GHz.  As noted above, the timescale sampled by the variability index varies between sources.  Nevertheless, there is no reason to expect the timescale sampled for a source to be related to its spectral properties and, consequently, the plot is still useful for investigating the relationship between spectral index and variability.  The plot shows a trend for sources with more steeply rising spectra to be increasingly variable.  \citet{bolton} have previously found evidence for a similar effect.  They found no evidence for variability in a group of sources with $\alpha > 0.5$, but 50~per~cent of sources with $\alpha < 0.5$ were found to be variable.

We have attempted to quantify the correlation between spectral index and variability, in our data, by calculating the mean spectral index for the sources defined as variable and non-variable at 16~GHz.  We find the mean spectral index for the variable sources as $-0.06 \pm 0.05$ and for the non-variable as $0.13 \pm 0.04$, where the uncertainties are standard errors on the means.  Given the level of the errors, we find a significant difference between the mean spectral indices of the two populations.

\section{Conclusions}\label{Conclusions}

In order to investigate the cm-wavelength-variability of flux density of sources selected at cm wavelengths, at the high-flux-density end of the source population, we have made observations with the AMI (13.9--18.2\:GHz) and VSA (33\:GHz) of a complete sample of sources found with \textit{WMAP} at 33\:GHz and with $\mathrm{S_{33\:GHz} \geq 1.1\:Jy}$.  We applied a new algorithm to data obtained by near-simultaneous -- typically a few minutes apart -- observations at our two observing bands.  This maximum-likelihood method allowed us to assess the correlation in the variability at the two frequencies, taking the measurement errors into account. We found that:

\begin{enumerate}

\item[(1)] on timescales of $\mathrm{\approx}$ 1.5~years, 15 per cent of sources varied by more than 20 per cent at 16~GHz and 20 per cent varied by more than 20 per cent at 33\:GHz;

\item[(2)] this level of variability means that variability has to be taken into account in coping with source contamination in CMB work, particularly at $\ell \gtrsim 2000$;

\item[(3)] this level of variability would indeed seem to explain the Eddington bias found in Paper I;

\item[(4)] in a subset of sources classified as variable from both the AMI and VSA data, the Pearson product-moment correlation coefficient between the variability at 16 and 33~GHz is $\mathrm{0.955 \pm 0.034}$ and the degree of variability at 16~GHz ($\mathrm{0.202 \pm 0.028}$) is very similar to that at 33\:GHz ($\mathrm{0.224 \pm 0.039}$);

\item[(5)] there is a significant difference between the mean spectral indices of the variable sources ($-0.06 \pm 0.05$) and non-variable sources ($0.13 \pm 0.04$);

\item[(6)] the AMI measurements show no strong evidence of a correlation between variability and flux density for the sample as a whole. There appears also to be little difference between the general levels of variability in our sample and the \citet{sadler} sample complete to 100\:mJy at 20\:GHz. These results seem somewhat surprising given that we do find a relation between spectral index and flux density.

\end{enumerate}

\section*{Acknowledgments}\label{acknowledgements}

We are grateful to the staff of the Mullard Radio Astronomy Observatory for the maintenance and operation of the AMI and to the staff of the Teide Observatory, Mullard Radio Astronomy Observatory and Jodrell Bank Observatory for assistance in the day-to-day operation of the VSA.  We are also grateful to the University of Cambridge and PPARC/STFC for funding and supporting the AMI.  We thank PPARC for funding and supporting the VSA project and the Spanish Ministry of Science and Technology for partial financial support (project AYA2001-1657).  MLD, TMOF, NHW, MO, CRG and TWS are grateful for support from PPARC/STFC studentships.

\begin{table*}
 \begin{minipage}{200mm}
 \caption{Results for individual sources.  The columns are the source name, taken from the NEWPS catalogue; the Equatorial \newline coordinates (J2000), from the PMN or GB6 catalogues; the number of observations, $n$; the mean flux density, $\bar{S}$; the source \newline class, variable (VAR) or non-variable (NV); the mean variability index, $V$, with the mean number of days separating the \newline observations shown in brackets and the spectral index, $\alpha_{13.9}^{33.75}$.}
 \label{tab:sources}
 \begin{tabular}{@{}l c c c c r r c c r@{} l r@{} l r@{} c@{} l}
 \hline
 Source & $\mathrm{\alpha (J2000)}$ & $\mathrm{\delta (J2000)}$ & \multicolumn{2}{|c|}{$n$} & \multicolumn{2}{|c|}{${\bar{S}}$/Jy} & \multicolumn{2}{|c|}{Class} & \multicolumn{4}{|c|}{$V$/per~cent} & \multicolumn{3}{|c|}{$\alpha_{13.9}^{33.75}$}\\
  &  &  & AMI & VSA & AMI & VSA & AMI & VSA & \multicolumn{2}{|c|}{AMI} & \multicolumn{2}{|c|}{VSA} & & & \\   
 \hline
 J0029$+$0554&00 29 45.9&$+$05 54 41&3&3& 0.65 &  0.66 &NV &NV &$ 1.1~$&(94) &$-1.0~$&(43) &$ 0.22~$&$\pm$&~0.19\\
 J0057$+$3021&00 57 48.3&$+$30 21 14&3&5& 0.79 &  0.96 &NV &NV &$ 0.9~$&(208)&$-8.1~$&(206)&$ 0.06~$&$\pm$&~0.16\\
 J0105$+$4819&01 05 50.8&$+$48 19 01&5&6& 0.49 &  0.58 &NV &NV &$ 1.2~$&(234)&$-0.1~$&(202)&$ 0.47~$&$\pm$&~0.16\\
 J0108$+$0134&01 08 38.7&$+$01 34 51&4&6& 1.50 &  1.37 &VAR&VAR&$14.8~$&(215)&$12.5~$&(236)&$-0.05~$&$\pm$&~0.12\\
 J0108$+$1319&01 08 52.7&$+$13 19 17&3&5& 1.60 &  1.00 &NV &NV &$-2.7~$&(277)&$-8.8~$&(227)&$     ~$&$ - $&~    \\
 J0136$+$4751&01 36 58.8&$+$47 51 27&4&4& 3.41 &  3.69 &VAR&VAR&$ 7.5~$&(320)&$ 5.6~$&(245)&$ 0.04~$&$\pm$&~0.07\\
 J0152$+$2206&01 52 17.8&$+$22 06 58&3&4& 0.88 &  0.95 &NV &NV &$-2.6~$&(289)&$-7.4~$&(245)&$ 0.12~$&$\pm$&~0.19\\
 J0204$+$1514&02 04 50.8&$+$15 14 10&4&5& 1.08 &  0.96 &VAR&NV &$ 8.6~$&(269)&$-1.2~$&(252)&$ 0.33~$&$\pm$&~0.17\\
 J0205$+$3212&02 05 04.1&$+$32 12 29&4&4& 2.76 &  2.98 &VAR&NV &$10.4~$&(320)&$-0.6~$&(244)&$-0.05~$&$\pm$&~0.08\\
 J0221$+$3556&02 21 05.8&$+$35 56 13&4&5& 1.27 &  1.13 &VAR&NV &$ 4.3~$&(228)&$ 6.3~$&(206)&$ 0.28~$&$\pm$&~0.14\\
 J0223$+$4259&02 23 14.5&$+$42 59 19&3&5& 0.94 &  1.19 &NV &NV &$ 2.7~$&(291)&$-3.5~$&(206)&$-0.06~$&$\pm$&~0.20\\
 J0237$+$2848&02 37 52.4&$+$28 48 14&4&4& 2.91 &  3.06 &VAR&VAR&$ 3.1~$&(269)&$ 5.9~$&(244)&$-0.14~$&$\pm$&~0.14\\
 J0238$+$1637&02 38 38.5&$+$16 37 04&5&4& 2.33 &  3.32 &VAR&VAR&$11.4~$&(234)&$10.2~$&(285)&$-0.28~$&$\pm$&~0.05\\
 J0303$+$4716&03 03 34.8&$+$47 16 19&4&4& 0.79 &  0.77 &VAR&NV &$ 8.4~$&(320)&$ 6.2~$&(244)&$ 0.00~$&$\pm$&~0.23\\
 J0319$+$4130&03 19 47.1&$+$41 30 42&4&4&14.93 & 11.51 &VAR&NV &$ 8.5~$&(320)&$ 1.7~$&(244)&$ 0.34~$&$\pm$&~0.06\\
 J0336$+$3218&03 36 30.0&$+$32 18 36&4&5& 0.92 &  1.01 &NV &NV &$ 1.4~$&(270)&$-6.3~$&(227)&$ 0.07~$&$\pm$&~0.15\\
 J0339$-$0146&03 39 30.4&$-$01 46 38&3&5& 1.92 &  1.89 &NV &NV &$ 0.6~$&(211)&$-5.1~$&(227)&$ 0.10~$&$\pm$&~0.13\\
 J0418$+$3801&04 18 22.4&$+$38 01 47&4&5& 5.41 &  7.20 &VAR&VAR&$16.7~$&(269)&$12.4~$&(227)&$-0.35~$&$\pm$&~0.06\\
 J0423$-$0120&04 23 15.8&$-$01 20 34&3&5& 4.44 &  4.85 &NV &NV &$-2.2~$&(211)&$-0.1~$&(227)&$-0.01~$&$\pm$&~0.08\\
 J0423$+$4150&04 23 55.7&$+$41 50 06&4&5& 1.32 &  1.25 &VAR&VAR&$ 7.4~$&(269)&$ 9.7~$&(226)&$ 0.32~$&$\pm$&~0.13\\
 J0424$+$0036&04 24 46.6&$+$00 36 05&1&1& 0.50 &  0.40 &$-$&$-$&$~~~-~$&~~~~~&$~~~-~$&~~~~~&$ 0.58~$&$\pm$&~0.26\\
 J0433$+$0521&04 33 11.0&$+$05 21 13&3&5& 2.95 &  2.82 &NV &VAR&$ 2.5~$&(207)&$ 6.9~$&(227)&$ 0.33~$&$\pm$&~0.09\\
 J0437$+$2940&04 37 04.7&$+$29 40 02&4&4& 4.47 &  2.45 &VAR&NV &$ 4.9~$&(269)&$-0.2~$&(261)&$ 0.94~$&$\pm$&~0.15\\
 J0449$+$1121&04 49 07.6&$+$11 21 25&2&4& 0.86 &  1.02 &VAR&VAR&$15.4~$&(319)&$14.7~$&(252)&$ 0.55~$&$\pm$&~0.28\\
 J0501$-$0159&05 01 12.9&$-$01 59 21&3&5& 0.92 &  0.99 &NV &NV &$-2.4~$&(211)&$-2.4~$&(212)&$ 0.16~$&$\pm$&~0.17\\
 J0533$+$4822&05 33 15.6&$+$48 22 59&3&4& 1.00 &  1.08 &VAR&NV &$11.6~$&(288)&$-0.1~$&(252)&$-0.13~$&$\pm$&~0.16\\
 J0555$+$3948&05 55 31.7&$+$39 48 45&3&4& 3.13 &  2.74 &NV &NV &$ 3.4~$&(293)&$ 2.0~$&(251)&$ 0.41~$&$\pm$&~0.13\\
 J0646$+$4451&06 46 31.4&$+$44 51 22&3&4& 3.20 &  2.76 &VAR&VAR&$ 5.9~$&(288)&$ 4.8~$&(252)&$ 0.21~$&$\pm$&~0.14\\
 J0733$+$5022&07 33 52.8&$+$50 22 18&4&5& 0.78 &  0.80 &VAR&NV &$11.6~$&(269)&$-5.1~$&(233)&$ 0.00~$&$\pm$&~0.14\\
 J0738$+$1742&07 38 07.6&$+$17 42 26&4&4& 0.78 &  0.80 &NV &NV &$ 0.2~$&(269)&$-4.1~$&(266)&$ 0.07~$&$\pm$&~0.16\\
 J0739$+$0137&07 39 18.2&$+$01 37 06&3&5& 1.26 &  1.80 &VAR&VAR&$21.8~$&(214)&$25.1~$&(233)&$ 0.09~$&$\pm$&~0.15\\
 J0750$+$1231&07 50 51.2&$+$12 31 13&3&4& 4.18 &  4.02 &NV &NV &$ 2.5~$&(214)&$ 1.0~$&(267)&$-0.01~$&$\pm$&~0.08\\
 J0757$+$0956&07 57 06.4&$+$09 56 21&3&4& 1.28 &  1.65 &NV &NV &$ 2.5~$&(208)&$-8.2~$&(266)&$-0.19~$&$\pm$&~0.13\\
 J0825$+$0309&08 25 49.5&$+$03 09 25&4&5& 1.30 &  1.89 &VAR&VAR&$13.6~$&(169)&$15.5~$&(232)&$-0.24~$&$\pm$&~0.13\\
 J0830$+$2410&08 30 52.3&$+$24 10 47&4&3& 1.11 &  1.28 &VAR&NV &$16.6~$&(269)&$ 8.1~$&(296)&$-0.09~$&$\pm$&~0.16\\
 J0840$+$1312&08 40 48.0&$+$13 12 37&4&3& 0.97 &  1.06 &VAR&NV &$ 4.6~$&(169)&$ 7.4~$&(295)&$ 0.21~$&$\pm$&~0.19\\
 J0854$+$2006&08 54 48.4&$+$20 06 47&3&3& 2.85 &  3.37 &VAR&NV &$20.7~$&(288)&$ 4.7~$&(296)&$-0.30~$&$\pm$&~0.09\\
 J0909$+$0121&09 09 09.5&$+$01 21 38&3&5& 1.21 &  1.32 &NV &NV &$ 3.7~$&(211)&$ 0.1~$&(232)&$ 0.08~$&$\pm$&~0.14\\
 J0920$+$4441&09 20 58.7&$+$44 41 44&3&4& 1.92 &  2.34 &NV &NV &$ 2.8~$&(289)&$-3.2~$&(251)&$-0.32~$&$\pm$&~0.12\\
 J0927$+$3902&09 27 03.0&$+$39 02 18&4&5& 9.75 &  8.49 &NV &NV &$-0.8~$&(269)&$-0.9~$&(233)&$ 0.19~$&$\pm$&~0.05\\
 J0948$+$4039&09 48 55.2&$+$40 39 56&3&4& 1.59 &  1.55 &VAR&NV &$ 9.2~$&(291)&$ 1.0~$&(252)&$-0.07~$&$\pm$&~0.15\\
 J0958$+$4725&09 58 19.9&$+$47 25 14&5&6& 1.26 &  1.16 &NV &NV &$ 2.2~$&(234)&$ 2.3~$&(206)&$ 0.23~$&$\pm$&~0.11\\
 J1033$+$4115&10 33 03.9&$+$41 15 59&2&3& 0.81 &  0.96 &NV &NV &$-2.9~$&(32) &$ 3.0~$&(43) &$-0.22~$&$\pm$&~0.18\\
 J1038$+$0512&10 38 47.7&$+$05 12 16&4&6& 1.41 &  1.21 &NV &NV &$ 2.4~$&(201)&$ 8.1~$&(207)&$ 0.40~$&$\pm$&~0.13\\
 J1041$+$0610&10 41 17.6&$+$06 10 02&3&7& 1.38 &  1.36 &NV &NV &$ 0.6~$&(177)&$-5.6~$&(223)&$ 0.07~$&$\pm$&~0.15\\
 J1058$+$0133&10 58 30.5&$+$01 33 46&3&6& 4.59 &  5.92 &NV &VAR&$ 1.9~$&(180)&$ 7.7~$&(207)&$-0.18~$&$\pm$&~0.06\\
 J1130$+$3815&11 30 54.6&$+$38 15 10&5&6& 1.35 &  1.40 &VAR&VAR&$ 3.2~$&(234)&$ 5.0~$&(206)&$-0.02~$&$\pm$&~0.10\\
 J1153$+$4931&11 53 24.7&$+$49 31 13&5&6& 1.20 &  1.29 &VAR&NV &$ 3.0~$&(234)&$-9.2~$&(206)&$ 0.01~$&$\pm$&~0.11\\
 J1159$+$2914&11 59 32.1&$+$29 14 53&4&5& 1.56 &  1.61 &VAR&VAR&$27.6~$&(269)&$28.6~$&(233)&$-0.20~$&$\pm$&~0.13\\
 J1219$+$0549&12 19 18.0&$+$05 49 39&0&4& $-$~~&  1.86 &$-$&NV &$~~~-~$&~~~~~&$-7.3~$&(251)&$ 0.80~$&$\pm$&~0.35\\
 J1229$+$0203&12 29 05.6&$+$02 03 09&3&5&29.20 & 26.98 &NV &NV &$-2.2~$&(185)&$ 0.1~$&(233)&$ 0.07~$&$\pm$&~0.05\\
 J1230$+$1223&12 30 48.8&$+$12 23 36&3&5&24.49 & 17.67 &NV &NV &$-2.8~$&(184)&$ 0.1~$&(233)&$ 0.41~$&$\pm$&~0.05\\
 J1310$+$3220&13 10 29.5&$+$32 20 51&4&5& 1.21 &  1.79 &VAR&VAR&$13.4~$&(228)&$12.8~$&(212)&$-0.61~$&$\pm$&~0.11\\
 J1331$+$3030&13 31 08.0&$+$30 30 35&4&5& 3.34 &  2.06 &NV &NV &$-2.7~$&(228)&$-0.1~$&(212)&$ 0.63~$&$\pm$&~0.10\\
 J1347$+$1217&13 47 33.4&$+$12 17 17&3&4& 1.36 &  1.03 &NV &NV &$-2.8~$&(270)&$ 4.1~$&(252)&$ 0.62~$&$\pm$&~0.20\\
 J1357$+$1919&13 57 04.1&$+$19 19 19&3&4& 1.98 &  2.36 &NV &NV &$ 1.5~$&(292)&$ 1.4~$&(251)&$-0.35~$&$\pm$&~0.13\\
 J1419$+$3822&14 19 45.9&$+$38 22 01&4&4& 0.60 &  0.86 &VAR&VAR&$41.7~$&(315)&$20.6~$&(252)&$-0.08~$&$\pm$&~0.18\\
 J1504$+$1029&15 04 24.0&$+$10 29 43&3&4& 1.60 &  1.50 &VAR&NV &$ 4.1~$&(270)&$-1.2~$&(251)&$ 0.00~$&$\pm$&~0.17\\
 J1516$+$0014&15 16 40.7&$+$00 14 57&3&4& 1.01 &  1.07 &NV &NV &$-2.7~$&(270)&$-4.8~$&(252)&$ 0.06~$&$\pm$&~0.22\\
 J1549$+$0237&15 49 30.0&$+$02 37 01&3&4& 2.16 &  2.19 &NV &NV &$-2.6~$&(270)&$-7.4~$&(251)&$ 0.08~$&$\pm$&~0.15\\
 J1550$+$0527&15 50 35.2&$+$05 27 06&3&4& 2.82 &  2.84 &NV &NV &$ 1.9~$&(270)&$ 1.4~$&(244)&$ 0.02~$&$\pm$&~0.10\\
 \end{tabular}								   				     
\end{minipage}								    				     
\end{table*}								    	
									    	
\begin{table*}								    	
\begin{minipage}{200mm}							    	
 \contcaption{}
 \begin{tabular}{@{}l c c c c r r c c r@{} l r@{} l r@{} c@{} l}	    	
 J1608$+$1029&16 08 46.4&$+$10 29 05&3&4&1.34& 1.44&NV &VAR&$ 0.6~$&(270)&$ 6.8~$&(244)&$ 0.11~$&$\pm$&~0.19\\
 J1613$+$3412&16 13 40.8&$+$34 12 41&4&4&3.21& 2.68&VAR&NV &$ 6.0~$&(315)&$ 5.3~$&(244)&$ 0.23~$&$\pm$&~0.08\\
 J1635$+$3808&16 35 15.6&$+$38 08 13&4&4&2.48& 2.58&VAR&NV &$ 5.9~$&(315)&$ 2.4~$&(244)&$-0.12~$&$\pm$&~0.12\\
 J1638$+$5720&16 38 13.0&$+$57 20 29&4&4&1.92& 2.45&VAR&NV &$ 2.6~$&(315)&$-1.2~$&(244)&$-0.34~$&$\pm$&~0.12\\
 J1642$+$3948&16 42 58.0&$+$39 48 42&4&4&5.30& 5.26&VAR&VAR&$ 9.5~$&(315)&$ 9.4~$&(244)&$ 0.18~$&$\pm$&~0.07\\
 J1651$+$0459&16 51 09.2&$+$04 59 33&3&4&2.63& 1.33&NV &NV &$-2.3~$&(270)&$-5.3~$&(244)&$ 1.03~$&$\pm$&~0.18\\
 J1720$-$0058&17 20 29.7&$-$00 58 37&0&5&$-$~~& 3.87&$-$&NV &$~~~-~$&     &$-3.9~$&(33) &$     ~$&$ - $&~    \\
 J1727$+$4530&17 27 28.4&$+$45 30 49&4&6&0.57& 0.94&VAR&VAR&$31.8~$&(308)&$29.8~$&(237)&$-0.67~$&$\pm$&~0.17\\
 J1734$+$3857&17 34 20.5&$+$38 57 45&4&4&0.89& 0.93&VAR&VAR&$ 7.0~$&(315)&$16.3~$&(244)&$-0.21~$&$\pm$&~0.17\\
 J1740$+$5211&17 40 36.6&$+$52 11 47&4&4&1.02& 0.95&VAR&VAR&$15.3~$&(314)&$13.5~$&(244)&$ 0.12~$&$\pm$&~0.17\\
 J1743$-$0350&17 43 59.2&$-$03 50 06&4&5&3.80& 3.66&VAR&VAR&$11.1~$&(245)&$11.9~$&(206)&$ 0.28~$&$\pm$&~0.07\\
 J1751$+$0938&17 51 32.7&$+$09 38 58&3&4&5.08& 6.36&VAR&VAR&$33.4~$&(270)&$ 7.2~$&(244)&$-0.19~$&$\pm$&~0.07\\
 J1753$+$2847&17 53 42.5&$+$28 47 58&3&4&1.78& 1.85&NV &NV &$ 1.7~$&(293)&$-0.9~$&(244)&$-0.14~$&$\pm$&~0.13\\
 J1801$+$4404&18 01 32.2&$+$44 04 09&4&4&1.40& 1.47&VAR&NV &$ 6.1~$&(314)&$ 7.5~$&(244)&$ 0.04~$&$\pm$&~0.15\\
 J1824$+$5650&18 24 06.8&$+$56 50 59&4&5&1.28& 1.34&VAR&NV &$11.3~$&(309)&$ 5.5~$&(226)&$-0.26~$&$\pm$&~0.12\\
 J1829$+$4844&18 29 32.1&$+$48 44 46&3&4&2.70& 2.32&NV &NV &$-2.2~$&(293)&$-5.8~$&(244)&$ 0.16~$&$\pm$&~0.13\\
 J1955$+$5131&19 55 42.3&$+$51 31 54&4&3&1.18& 1.50&VAR&NV &$16.5~$&(314)&$-8.4~$&(227)&$-0.63~$&$\pm$&~0.14\\
 J1959$+$4034&19 59 21.8&$+$40 34 28&0&3&$-$~~&37.43&$-$&NV &$~~~-~$&     &$-1.8~$&(319)&$     ~$&$ - $&~    \\
 J2123$+$0535&21 23 43.4&$+$05 35 14&3&3&1.41& 1.41&NV &NV &$ 4.9~$&(278)&$-6.3~$&(229)&$ 0.00~$&$\pm$&~0.19\\
 J2134$-$0153&21 34 10.4&$-$01 53 25&3&2&1.90& 1.75&NV &NV &$-2.4~$&(277)&$-3.1~$&(334)&$ 0.27~$&$\pm$&~0.19\\
 J2136$+$0041&21 36 38.6&$+$00 41 54&3&6&5.76& 4.22&NV &NV &$-2.6~$&(183)&$ 1.3~$&(183)&$ 0.42~$&$\pm$&~0.05\\
 J2139$+$1423&21 39 01.5&$+$14 23 37&3&4&2.13& 1.75&NV &NV &$ 3.2~$&(221)&$-7.3~$&(181)&$ 0.33~$&$\pm$&~0.08\\
 J2143$+$1743&21 43 35.6&$+$17 43 54&4&6&0.50& 0.57&VAR&NV &$12.9~$&(206)&$-8.0~$&(201)&$-0.23~$&$\pm$&~0.14\\
 J2148$+$0657&21 48 05.5&$+$06 57 36&3&5&5.52& 5.07&NV &VAR&$ 2.0~$&(212)&$ 4.3~$&(206)&$ 0.19~$&$\pm$&~0.05\\
 J2202$+$4216&22 02 44.3&$+$42 16 39&4&5&3.47& 4.33&VAR&VAR&$12.0~$&(228)&$13.3~$&(207)&$-0.26~$&$\pm$&~0.05\\
 J2203$+$1725&22 03 26.7&$+$17 25 42&4&5&1.18& 1.25&VAR&VAR&$10.3~$&(228)&$ 7.8~$&(206)&$-0.07~$&$\pm$&~0.11\\
 J2203$+$3145&22 03 15.8&$+$31 45 38&4&5&2.62& 2.60&NV &NV &$-0.7~$&(228)&$ 0.7~$&(206)&$-0.08~$&$\pm$&~0.08\\
 J2212$+$2355&22 12 05.9&$+$23 55 31&4&4&0.95& 0.96&VAR&NV &$ 3.5~$&(228)&$ 3.6~$&(179)&$-0.14~$&$\pm$&~0.12\\
 J2218$-$0335&22 18 51.8&$-$03 35 40&3&4&1.65& 1.33&NV &NV &$ 1.8~$&(213)&$-3.7~$&(181)&$ 0.47~$&$\pm$&~0.15\\
 J2225$+$2118&22 25 37.6&$+$21 18 17&4&4&1.11& 1.27&NV &NV &$ 0.1~$&(228)&$ 2.3~$&(179)&$-0.07~$&$\pm$&~0.11\\
 J2232$+$1143&22 32 36.6&$+$11 43 54&4&4&3.79& 4.52&VAR&VAR&$ 5.7~$&(252)&$ 3.3~$&(180)&$-0.29~$&$\pm$&~0.05\\
 J2236$+$2828&22 36 20.8&$+$28 28 56&5&5&0.99& 1.60&VAR&VAR&$12.6~$&(234)&$ 8.3~$&(168)&$-0.51~$&$\pm$&~0.10\\
 J2253$+$1608&22 53 58.0&$+$16 08 53&4&5&7.17&10.73&VAR&VAR&$14.8~$&(228)&$22.8~$&(206)&$-0.56~$&$\pm$&~0.05\\
 J2327$+$0940&23 27 33.1&$+$09 40 02&4&6&1.81& 2.18&VAR&VAR&$ 6.8~$&(201)&$12.8~$&(201)&$-0.27~$&$\pm$&~0.08\\
 J2335$-$0131&23 35 20.1&$-$01 31 14&3&6&0.56& 0.79&VAR&NV &$15.3~$&(184)&$-0.5~$&(183)&$ 0.21~$&$\pm$&~0.17\\
 J2354$+$4553&23 54 21.9&$+$45 53 00&4&5&0.98& 0.77&NV &NV &$-1.7~$&(228)&$-7.4~$&(206)&$ 0.32~$&$\pm$&~0.14\\
  \hline
 \end{tabular}
\end{minipage}
\end{table*}
\label{lastpage}

\begin{thebibliography}{}

\bibitem[\protect\citeauthoryear{Akritas \& Bershady}{1996}]{akritas}
Akritas~M.~G., Bershady~M.~A., 1996, ApJ, 470, 706 

\bibitem[\protect\citeauthoryear{Barvainis et al.}{2005}]{barvainis}
Barvainis~R., Leh\'{a}r~J., Birkinshaw~M., Falcke~H., Blundell~K.~M., 2005, ApJ, 618, 108

\bibitem[\protect\citeauthoryear{Bennett et al.}{2003}]{bennett}
Bennett~C.~L., Bay~M., Halpern~M. et al., 2003, ApJ, 583, 1

\bibitem[\protect\citeauthoryear{Blandford \& Konigl}{1979}]{blandford}
Blandford~R.~D., Konigl~A., 1979, ApJ, 232, 34  

\bibitem[\protect\citeauthoryear{Bock, Large \& Sadler}{1999}]{bock}
Bock~D.~C.-J., Large~M.~I., Sadler~E.~M., 1999, AJ, 117, 1578

\bibitem[\protect\citeauthoryear{Bolton et al.}{2006}]{bolton}
Bolton~R.~C., Chandler~C.~J., Cotter~G., Pearson~T.~J., Pooley~G.~G., Readhead~A.~C.~S., Riley~J.~M., Waldram~E.~M., 2006, MNRAS, 370, 1556 

\bibitem[\protect\citeauthoryear{Cawthorne \& Rickett}{1985}]{cawthorne}
Cawthorne~T.~V., Rickett~B.~J., 1985, Nat, 315, 40 

\bibitem[\protect\citeauthoryear{Condon et al.}{1998}]{condon}
Condon~J.~J., Cotton~W.~D., Greisen~E.~W., Yin~Q.~F., Perley~R.~A., Taylor~G.~B., Broderick~J.~J., 1998, AJ, 115, 1693

\bibitem[AMI Consortium: Davies et al. (2009)]{davies}
Davies~M.~L., Franzen~T.~M.~O., Davis~R.~J. et al., 2009, MNRAS, in press

\bibitem[\protect\citeauthoryear{de~Zotti et al.}{2005}]{zotti}
de~Zotti~G., Ricci~R., Mesa~D., Silva~L., Mazzotta~P., Toffolatti~L., Gonz{\'a}lez-Nuevo~J., 2005, A\&A, 431, 893

\bibitem[\protect\citeauthoryear{Dennett-Thorpe \& de~Bruyn}{2002}]{dennett-thorpe}
Dennett-Thorpe~J., de~Bruyn~A.~G., 2002, Nat, 415, 57 

\bibitem[\protect\citeauthoryear{Dent}{1965}]{dent}
Dent~W.~A., 1965, Sci, 148, 1458

\bibitem[\protect\citeauthoryear{Feroz \& Hobson}{2008}]{feroz}
Feroz~F., Hobson~M.~P., 2008, MNRAS, 384, 449

\bibitem[\protect\citeauthoryear{Feroz, Hobson \& Bridges}{2008}]{feroz2}
Feroz~F., Hobson~M.~P., Bridges~M., 2008, preprint (arXiv:0809.3437)

\bibitem[\protect\citeauthoryear{Gonz{\'a}lez-Nuevo et al.}{2006}]{gonzalez-nuevo06}
Gonz{\'a}lez-Nuevo~J., Arg{\"u}eso~F., L{\'o}pez-Caniego~M., Toffolatti~L., Sanz~J.~L., Vielva~P., Herranz~D., 2006, MNRAS, 369, 1603

\bibitem[\protect\citeauthoryear{Gregorini, Ficarra \& Padrielli}{1986}]{gregorini}
Gregorini~L., Ficarra~A., Padrielli~L., 1986, A\&A, 168, 25 

\bibitem[\protect\citeauthoryear{Gregory et al.}{1996}]{gregory}
Gregory~P.~C., Scott~W.~K., Douglas~K., Condon~J.~J., 1996, ApJS, 103, 427

\bibitem[\protect\citeauthoryear{Griffith \& Wright}{1993}]{griffith}
Griffith~M.~R., Wright~A.~E., 1993, AJ, 105, 1666

\bibitem[\protect\citeauthoryear{Hinshaw et al.}{2007}]{hinshaw}
Hinshaw~G., Nolta~M.~R., Bennett~C.~L. at al., 2007, ApJS, 170, 288

\bibitem[\protect\citeauthoryear{Hirabayashi et al.}{2000}]{hirabayashi}
Hirabayashi~H., Fomalont~E.~B., Horiuchi~S. et al., PASJ, 2000, 52, 997

\bibitem[\protect\citeauthoryear{Jauncey et al.}{2000}]{jauncey}
Jauncey~D.~L., Kedziora-Chudczer~L.~L., Lovell~J.~E.~J., Nicolson~G.~D., Perley~R.~A., Reynolds~J.~E., Tzioumis~A.~K., Wieringa~M.~H., 2000, in Hirabayashi~H., Edwards~P~G., Murphy~D.~W., eds., Proceedings of the VSOP Symposium, Astrophysical Phenomena Revealed by Space VLBI, p.147

\bibitem[\protect\citeauthoryear{Lang \& Rickett}{1970}]{lang}
Lang~K.~R., Rickett~B.~J., 1970, Nat, 225, 528 

\bibitem[\protect\citeauthoryear{Linsky, Rickett \& Redfield}{2008}]{linsky}
Linsky~J.~L., Rickett~B.~J., Redfield~S., 2008, ApJ, 675, 413 

\bibitem[\protect\citeauthoryear{L{\'o}pez-Caniego et al.}{2007}]{lopez-caniego}
L{\'o}pez-Caniego~M., Gonz{\'a}lez-Nuevo~J., Herranz~D., Massardi~M., Sanz~J.~L., de~Zotti~G., Toffolatti~L., Arg\"ueso~F., 2007, ApJS, 170, 108

\bibitem[\protect\citeauthoryear{Owen, Spanger \& Cotton}{1980}]{owen}
Owen~F.~N., Spanger~S.~R., Cotton~W.~D., 1980, AJ, 85, 351 

\bibitem[\protect\citeauthoryear{Rickett, Coles \& Bourgois}{1984}]{rickett1984}
Rickett~B.~J., Coles~W.~A., Bourgois~G., 1984, A\&A, 134, 390 

\bibitem[\protect\citeauthoryear{Rickett, Lazio \& Ghigo}{2006}]{rickett2006}
Rickett~B.~J., Lazio~T.~J.~W., Ghigo~F.~D., 2006, ApJS, 165, 439 

\bibitem[\protect\citeauthoryear{Sadler et al.}{2006}]{sadler}
Sadler E.~M., Ricci~R., Ekers~R.~D. et al., 2006, MNRAS, 371, 898 

\bibitem[\protect\citeauthoryear{Shapirovskaya}{1978}]{shapirovskaya}
Shapirovskaya~N.~Y., 1978, SvA, 22, 544 

\bibitem[\protect\citeauthoryear{Spangler et al.}{1989}]{spangler}
Spangler~S., Fanti~R., Gregorini~L., Padrielli~L., 1989, A\&A, 209, 315 

\bibitem[\protect\citeauthoryear{Tingay et al.}{2003}]{tingay}
Tingay~S.~J., Jauncey~D.~L., King~E.~A., Tzioumis~A.~K., Lovell~J.~E.~J., Edwards~P.~G., 2003, PASJ, 55, 351 

\bibitem[\protect\citeauthoryear{Toffolatti et al.}{1998}]{toffolatti}
Toffolatti~L., Arg\"ueso~G\'{o}mez~F., de~Zotti~G., Mazzei~P.~F., Franceschini~A., Danese~L., Burigana~C., 1998, MNRAS, 297, 117

\bibitem[\protect\citeauthoryear{Watson et al.}{2003}]{watson}
Watson~R.~A., Carreira~P., Cleary~K. et al., 2003, MNRAS, 341, 1057

\bibitem[\protect\citeauthoryear{Wright et al.}{2009}]{wright}
Wright~E.~L., Chen~X., Odegard~N. et al., 2009, ApJS, 180, 283

\bibitem[\protect\citeauthoryear{Zensus \& Pearson}{1987}]{zensus}
Zensus~J.~A., Pearson~T.~J., 1987, Journal of the British Astronomical Association, 98, 48 

\bibitem[\protect\citeauthoryear{AMI Consortium: Zwart et al.}{2008}]{zwart}
Zwart~J.~T.~L., Barker~R.~W., Biddulph~P. et al., 2008, MNRAS, 391, 1545

\end{thebibliography}
\end{document}